# Supra-operonic clusters of functionally related genes (SOCs) are a source of horizontal gene co-transfers

Tin Yau Pang[1,*] and Martin J. Lercher[1]

[1] Institute for Computer Science, Heinrich Heine University, Düsseldorf, 40225, Germany

* Corresponding author
Email: pang@hhu.de

## ABSTRACT

Adaptation of bacteria occurs predominantly via horizontal gene transfer (HGT). While it is widely recognized that horizontal acquisitions frequently encompass multiple genes, it is unclear what the size distribution of successfully transferred DNA segments looks like and what evolutionary forces shape this distribution. Here, we identified 1790 gene family pairs that were consistently co-gained on the same branches across a phylogeny of 53 E. coli strains. We estimated a lower limit of their genomic distances at the time they were transferred to their host genomes; this distribution shows a sharp upper bound at 30 kb. The same gene-pairs can have larger distances (up to 70 kb) in other genomes. These more distant pairs likely represent recent acquisitions via transduction that involve the co-transfer of excised prophage genes, as they are almost always associated with intervening phage-associated genes. The observed distribution of genomic distances of co-transferred genes is much broader than expected from a model based on the co-transfer of genes within operons; instead, this distribution is highly consistent with the size distribution of supra-operonic clusters (SOCs), groups of co-occurring and co-functioning genes that extend beyond operons. Thus, we propose that SOCs form a basic unit of horizontal gene transfer.

## INTRODUCTION

Bacterial adaptation to changes in the environment often occurs through horizontal gene transfer (HGT)[1,2], i.e., the uptake of genes from genomes of other strains or even other species. HGT massively accelerates the spread of antibiotic drug resistance in bacteria[3]; major functions of horizontally transferred genes further include metabolism, DNA transformation, pathogenesis, toxin production[4]. Bacteria can exchange DNA through diverse mechanisms including transformation, transduction, conjugation, gene transfer agents, and nanotubes[5,6]. If the incoming DNA sequence is highly similar to sequences of the recipient bacterium, then it can be integrated via homologous recombination[7]. Otherwise, the foreign DNA segments may be added to the genome through non-homologous recombination after entering the host, resulting in HGT. If transferred genes confer phenotypic changes that provide fitness advantages, then they are likely to become fixed in the bacterial population. HGT is an inseparable aspect of bacterial evolution. On an evolutionary time scale, genes have frequently been transferred across species, generating a complex transfer network[8].

Bacterial genomes are highly dynamical[2]. In addition to gene acquisitions via HGT, genes are frequently lost via mutational deletions, a process accelerated by a mutational bias towards gene deletions[9]. Genes no longer required in the current environment(s) will thus eventually get lost from bacterial genomes. The local pan-genome, the union of all genes in the environment, can be viewed as a toolbox of genes, and HGT facilitates the acquisition of genes needed for adaptation from this toolbox[1,10,11]. Many phenotypes require the cooperation of two or more genes; accordingly, the joint presence or absence of two genes across many genomes can be used to identify functional associations between them, a method termed phylogenetic profiling[12].

Numerous comparative analyses of prokaryotic genome organization provide evidence for the conservation of genomic clusters beyond operons[13–16]. The emergence of supra-operonic clusters (SOCs) may be driven by organizational principles of bacterial genomes. Natural selection may favor the formation of SOCs by minimizing the distances of co-functioning genes or co-functioning operons[17,18]; as the chromosome has a compact structure, it needs to be unraveled to enable transcription, and hence co-functioning genes may be driven into the same DNA coil to minimize such unraveling[18]. As the rate of HGT is extremely slow on human timescales, the role of HGT in the emergence of such genomic clusters is hard to quantify directly. Instead, one must rely on models to test how different driving forces shape genomic clusters[15–17,19], and the size distribution of the clusters can serve as a simple and convenient criterion for distinguishing between models.

In this study, we revisit the idea of HGT as a driving force towards genomic clustering. HGT may not be responsible for operon formation[20]. However, operon structures concentrate sets of co-functioning genes on a relatively small continuous stretch of DNA, thereby facilitating their co-transfer and hence increasing the probability that any HGT event will be selectively favorable for the recipient[19]. Previous work has confirmed the functional and genomic clustering of co-gained gene-pairs: co-transferred gene-pairs in proteobacteria are indeed five times more likely to function in the same pathway compared to separately transferred genes[21]. The same study also found that co-transferred gene-pairs are more than twice as likely as random pairs to be genomic neighbours.

Further, it is often assumed that operons are the basic unit of HGT. We hypothesized that larger units, the SOCs, may also contribute to the co-transfer of interacting genes. To test if SOCs are basic units of HGT, we reconstructed the phylogenetic tree of 53 *E. coli* and *Shigella* strains, and identified gene-pairs that were consistently co-gained along the branches of this phylogeny. We found that *E. coli* operons are too small to explain the observed distance distribution of co-transferred genes. On the other hand, expectations from SOCs, defined based either on the co-occurrence or on the co-functioning of genes, closely match the empirical distance distribution. We conclude that SOCs form a basic unit of HGT in *E. coli*, and that successful co-transfers of genes are not predominantly constrained by the carrying capacities of transfer agents, but by the size distribution of SOCs.

## RESULTS

We identified orthologous gene families across 53 *E. coli* and *Shigella* strains (along with 17 strains of other species that served as the outgroup; Supplementary Table S1). *Shigella* strains are generally considered to belong to the species *E. coli*[22]; thus, we will subsume all 53 strains under the species name *E. coli* in the remainder of this paper. These 53 strains cover the major recognized clades of *E. coli* (A, B1, B2, D, and E); they thus form a representative sample of the *E. coli* species. Our sample of *E. coli* genomes includes several pairs of closely related strains, such as K-12 MG1655 and K-12 W3110. For each pair, we identified several differences in gene content (Supplementary Table S7); their recent origin provides high confidence in our inference of HGT and the estimated distance of co-transferred genes (see below).

We reconstructed a maximum-likelihood phylogeny based on the concatenated alignment of 1334 1-to-1 orthologs universally present in all 70 genomes. It has previously been shown that in maximum likelihood analyses of concatenated alignments, HGT influences branch lengths but hardly affects the topology of the resulting tree[23,24]. The resulting rooted tree topology thus represents vertical inheritance among the 53 *E. coli* strains. Each internal branch was retrieved in at least 60% of bootstrap samples (see Supplementary Fig. S1 for the *E. coli* tree; Supplementary Table S7 for the bootstrap values, gene gains, and gene losses on different branches; and Supplementary Fig. S2 for the tree including the outgroup strains). This confidence level is sufficient for the following inference of HGT instances; note that false HGT inference caused by a wrong tree topology will be biased against a systematic distance distribution of apparently co-transferred genes, and will thus only add noise to our analysis.

We considered each internal node of the phylogenetic tree as an ancestral genome. We reconstructed gene presence and absence in theses ancestral genomes using the maximum-likelihood algorithm implemented in GLOOME Computational analyses of transcriptomic data reveal the dynamic organization of the Escherichia coli chromosome under different conditions[25], which allows for gene-specific and branch-specific gene gain and loss rates. We considered a gene to be present in an ancestral genome if GLOOME assigned the gene to the corresponding node with a probability $P≥0.5$; otherwise, we considered the gene absent. A change from absent to present between two consecutive nodes corresponds to gene gain via HGT, the reverse change corresponds to a gene loss. To define a second, high-confidence set of horizontally transferred genes, we used a more conservative cutoff. A horizontally transferred gene is included in this high-confidence set if it is absent with probability ≥0.8 (*i.e.,* $P≤0.2$) in the ancestor node of a branch, and present with $P≥0.8$ in the descendant node. We report results for the more inclusive data set in the main text, and for the smaller, high-confidence data set in Supplementary Figures; the results of both sets are highly consistent with each other.

**Statistical association of gene family pairs across transfer events**

For each pair of orthologous gene families in our dataset, we calculated the score of co-gains of the two gene families across the 104 branches of the *E. coli* tree (see Methods for details). Gene family pairs that are co-gained more often than expected by chance indicate a functional co-operation of the genes. While many co-gained pairs likely occur through the simultaneous acquisition of two genes on one DNA segment, some co-gained pairs could also stem from distinct HGT events. We compared the distribution of co-gain scores for all gene family pairs in the empirical data with that of a null model based on randomizations. The score distributions for the empirical data and the random null model are significantly different (Fig. 1), indicating that some gene family pairs indeed show many more co-gains than expected by chance. The score distribution of co-gains was also calculated based on the high-confidence HGT set; the distributions are highly consistent with those based on all genes (Supplementary Fig. S3).

The false discovery rate (FDR) is the fraction of pairs at a given co-gain score $t$ for which this score is likely due to chance alone. It can be calculated as the number of pairs showing an equal or stronger association than $t$ in the null model, divided by the corresponding number for the empirical data. For all transferred genes, we examined co-gained orthologous family pairs at FDR 0.05, corresponding to a co-gain score of $t$=-7.1289, and at FDR 0.005, corresponding to a score of $t$=-9.1090 (see the two vertical dotted lines in Fig. 1). Table 1 shows the number of significantly co-gained gene family pairs at different FDR. For the genes in the high-confidence HGT set, the co-gain score corresponding to FDR 0.05 and 0.005 are $t$=-7.0039 and $t$=-9.1090, respectively (Supplementary Fig. S3).

**Distance distribution between co-gained orthologous gene family pairs**

In many cases, parts of horizontally transferred DNA segments will be lost immediately, thereby reducing the distance between two co-gained genes. Conversely, later genomic rearrangements may increase the original genomic distances. To estimate the genomic distance during the HGT event, we took the minimum distance between members of the two co-gained gene families in any of the examined genomes. If any of the genes in the pair has paralogous alleles present in a genome, we used the minimum distance of all possible allele pairs to represent the distance of the gene pair for this genome. Fig. 2 shows the cumulative distribution of the genomic distances between co-gained gene family pairs at different FDRs. All curves show a prominent kink at 30 kb. When considering all co-gained gene family pairs, we see an additional tail that extends from 30 kb up to around 2.5Mb (half of *E. coli's* genome size), while the exclusion of phage-associated gene families leads to a sharp cutoff at 30 kb. Very similar features are observed if the distribution curves are calculated based on the high-confidence HGT set (Supplementary Fig. S4). The tail caused by the inclusion of phage-associated gene family pairs indicates the existence of two stacked distributions in this dataset; the tail distribution likely reflects independent HGT events caused by transduction, as it is very similar to the distance distribution of randomly chosen gene-pairs in an *E. coli* genome (Supplementary Fig. S6, averaged over five different *E. coli* strains).

The dominant distribution at lower genomic distances to the left of the kink likely represents the pairwise distances of genes that were co-gained in a single HGT event; thus, 30 kb appears to be an upper bound on the size of successfully transferred DNA segments. To further explore the nature of this 30 kb kink, we plotted the pairwise distance distribution that treats each occurrence of a gene family pair on different extant genomes equally, instead of only considering the minimal distance of gene family members (Supplementary Fig. S7 for all co-gained pairs, Supplementary Fig. S8 for high-confidence pairs). These curves largely resemble the distributions in Fig. 2, except that a second kink emerges at 70 kb for all distance distributions.

An upper limit on the size of transferred DNA segments may be caused by the limited carrying capacity of transfer agents. The main mechanisms of HGT in *E. coli* are transduction and conjugation[26,27]. The majority (58%) of the phage genomes in the EMBL database[28] have a total sequence length between 30 kb and 70 kb (Supplementary Fig. S9), consistent with the hypothesis that the 70 kb kink in Supplementary Fig. S7 is caused by recently integrated DNA segments that reflect the limited size of phage capsids. Our observations are also consistent with a recent study on domesticated prophages[29], which reports a bimodal length distribution for prophages integrated into enterobacterial genomes; the short-range and long-range modes of this bimodal distribution are separated at 30 kb, while the long-range mode is bounded above at 70 kb. Bobay *et al.* suggest that the integrated phages in the long-range mode tend to be recently integrated prophages. As bacteria tend to lose DNA that does not contribute to fitness[9], these decay rapidly, shrinking to the short-range mode of the distribution which is then maintained by purifying selection[29].

This suggests that the majority of allele pairs that belong to co-gained gene families but which are located between 30 kb and 70 kb apart reflect recent HGT events through transduction. To test if the more distant gene-pairs indeed show evidence of phage-association, we compared close (distance<30 kb) and distant (30kb≤distance<70kb) allele pairs from co-gained non-phage-associated orthologous gene families (at FDR 0.05), assessing the percentage of intervening phage-associated-genes. As expected, we found that the average percentage of intervening phage-associated-genes (excluding those instances having no genes in between) is substantially larger for distant alleles (63.02%) than for close alleles (0.50%; $p<10^{-6}$, Wilcoxon rank sum test). Thus, co-gained allele pairs >30 kb apart likely represent recent acquisitions through phages.

Further analysis found that 89.27% of these distant allele pairs have at least one intervening phage-associated-gene; moreover, the maximum distance of the allele pairs within the remaining 10.73% is 35,418 bp, i.e., they are likely to be in the tail of the first mode of the distribution (close allele pairs). It is widely believe that there are two types of transduction: (i) random DNA segments incorporated during cell lysis, and (ii) flanking DNA of a prophage incorporated due to imprecise chromosome excision[2]. Our result seems to suggest that all recently integrated DNA segments belong to type (ii) transduction that involves the co-transfer of phage genes, while type (i) phage DNA-free transduction does not contribute to the observed transfer events.

The co-transferred prophage genes in the DNA between distant co-gained genes are not stable and will likely be lost over evolutionary time, causing the distance between the alleles of the co-gained gene-pair to shrink below 30 kb. In the remainder of this paper, we will not consider recent acquisitions, but focus on the distance distribution of non-phage-associated stable gene-pairs likely maintained by stabilizing selection (brown lines in Fig. 2).

## Operons cannot explain the distance distribution of co-gained gene family pairs

Which mechanistic or selective forces determine the shape of the distribution observed for non-phage-associated gene family pairs in Fig. 2? We argued above (as did others[29]) that this distance distribution reflects gene-pairs maintained by stabilizing selection; it thus reflects the genomic organization of functional relationships. Operons are the predominant unit of co-functional genes in bacterial genome organisation, and hence the distribution in Fig. 2 may simply reflect gene distances in operons; it has even been suggested that natural selection favours the organization of functionally related genes into "selfish" operons to facility HGT[19].

To test if operons may be responsible for the distance distribution of co-gained gene family pairs in Fig. 2, we compared this distribution with the pairwise distance distribution of genes in operons of five different *E. coli* strains. As can be seen from Fig. 3, the operon model distribution (black dashed line) is unable to fit the empirical distribution (brown): the distances between gene-pairs within *E. coli* operons are too short to explain the distance distribution of the observed pairs of co-gained gene families.

One might of course hypothesize that genes in a co-gained gene-pairs were originally in the same operon but increased their distance due to subsequent genomic rearrangements. However, such rearrangements would be inconsistent with the observed 30 kb cutoff.

## SOCs can be inferred from autocovariance of mutual conservation or gene co-functioning

Thus, to explain the distance distribution of consistently co-gained gene family pairs, we have to look at structures of functionally coupled genes that extend beyond operons, which we denote supra-operonic clusters (SOCs). Such larger functional genomic units have been reported previously[13–18], but are still poorly characterized. SOCs were first inferred from the conservation of gene clusters between species[14]. Accordingly, we first approximated the pairwise distance distribution of genes in SOCs through the normalized autocovariance of co-occurring gene-pairs (CO-AC), calculated from gene presence and absence across 233 γ-proteobacterial species including *E. coli* (see Supplementary Fig. S11 for the CO-AC curves at different cutoffs of mutual information (MI)). Genes with phage and mobile element association may be genomically clustered due to reasons unrelated to their cellular functions, and we thus excluded these from the CO-AC calculation.

Fig. 3 shows the cumulative distance distribution of gene-pairs in SOCs calculated via CO-AC (black dashed line). As expected from previous observations[13–16], we find that SOCs extend to much larger distances than operons (Fig. 3): while the median of gene distances in operons is 1.7kb, it is increased to 9.2kb in SOCs.

Genes within SOCs are expected to be co-functioning[13–18]. As an alternative to CO-AC, we also approximated the pairwise distance distribution of genes in SOCs through the normalized autocovariance of functionally similar gene-pairs (GO-AC), considering two genes to be functionally similar if they had at least one GO term[30] in common. Again, we excluded genes with phage and mobile element association. Fig. 3 shows the average of our calculation of the rescaled GO-AC for five *E. coli* strains (black dash-dot line). Reassuringly, the estimates based on co-occurrence (CO-AC) and on co-functioning (GO-AC) are highly consistent with each other (Fig. 3), despite their fundamentally different approaches to identify genes in SOCs.

To validate our estimation strategy for the distance distribution of gene-pairs in SOCs, we also calculated GO-AC restricted to gene-pairs within the same *E. coli* operon[31] (Supplementary Fig. S12). As expected, the operon-specific GO-AC is very similar to the pairwise distance distributions of genes within operons (Supplementary Fig. S12), confirming the validity of our approach.

To test how the distance distribution of gene-pairs depends on the level of co-functioning, we used the number of common GO terms between a pair of genes to measure their level of co-functionality, and calculated GO-$AC_n$ at different cutoffs $n$ for the number of common GO terms (Supplementary Fig. S13). As expected, increasing cutoffs lead to a faster decay of the GO-$AC_n$ curve to the background level; for large $n$, the curve approaches that of GO-$AC_1$ restricted to gene-pairs within operons. A similar effect is seen when requiring increasingly higher levels of co-occurrence. Thus, at more stringent cutoffs for both co-functioning and co-occurrence, the distribution is shifted to smaller distances until it resembles the distance distribution of gene-pairs within operons (Supplementary Fig. S14 for distance distributions averaged over 5 different *E. coli* strains at different $n$, and Supplementary Fig. S15 for the distance distributions of 5 different *E. coli* strains at $n$=1).

We expect the level of co-occurrence to be related to the level of co-functioning, i.e., frequently co-occurring gene-pairs should be more likely to be functionally related than less frequently co-occurring pairs. Supplementary Fig. S16 shows the boxplot of MI for categories of $n$>0; we indeed find a weak positive correlation between the number of shared GO categories and the mutual information from co-occurrence (Spearman's ρ=0.076, $p<10^{-15}$).

**SOCs can explain the distance distribution of co-gained genes**

The distance distribution of consistently co-gained gene-pairs (Fig. 3, brown solid line) is highly consistent with the distance distribution in SOCs estimated either from co-functioning (GO-AC, black dash-dot line) or from co-occurrence (CO-AC, black solid line). In particular, the distance distribution for gene-pairs in SOCs also drops to zero at around 30 kb (this is also the case when calculating CO-AC from γ-proteobacteria excluding *E. coli*, Supplementary Fig. S17 and S15).

## DISCUSSION

We identified consistently co-gained gene family pairs among 53 strains of *E. coli*. Excluding phage-associated-genes, we found that the minimal distance of these co-gained gene-pairs across the *E. coli* strains follows a distribution that is bounded above at 30 kb, indicating that 30 kb represents a natural upper bound for HGT. When considering all genomic distances within two consistently co-gained gene family pairs, a second mode bounded above at 70 kb appears. Gene-pairs in this long-distance mode often have intervening phage-associated genes and likely represent recent acquisitions of DNA segments mediated by phages; these recent acquisitions are expected to contract over evolutionary time. Our analysis reveals that transferred segments >30 kb are unlikely to be random DNA segments from other bacteria, but instead represent flanking DNA of prophage genes incorporated into the phage capsid with the prophage due to imprecise excision. A viral capsid can transfer a DNA segment as long as 70 kb, consistent with the observed distance distribution of co-transferred gene families across all genomes. While the phage genes will decay over evolutionary time scales, the accompanying gene sets, in case they are co-functioning and promoted by natural selection, can be retained at lengths up to 30 kb.

It is often assumed that operons form the basic unit of HGT[19]. However, we found that pairwise gene distances within operons are inconsistent with the observed distance distribution of co-gained gene families. Instead, the distance distribution of co-gained gene families is highly consistent with the distance distribution of genes within SOCs (Fig. 3), defined as sets of co-functioning and/or frequently co-occurring genes[14].

We estimated the pairwise distance distribution of genes in SOCs from the autocovariance (AC) of co-occurring (CO-AC) and of co-functioning (GO-AC) gene-pairs. Both CO-AC and GO-AC have a tunable cutoff, which allows one to probe the cluster structure of genes at different degrees of coherence. At high cutoff values (CO-cutoff=0.8, GO-cutoff=5), the curves likely reflect the strong coherence of gene-pairs located in the same operon. At relaxed cutoff values (CO-cutoff=0.0001, GO-cutoff=1), the curves likely represent the distribution of loosely connected clusters – SOCs – which extend up to 30 kb; further reduction of this cutoff value does not lead to a broader distribution. Thus, we propose that SOCs, rather than operons, might be considered the basic unit of HGT.

Previous work has linked SOCs to co-regulated operons, regional variation in codon bias, and chromosomal coiling[14,32–36]. The 30 kb cutoff also loosely coincides with the length scale of DNA supercoiling domains (average length ~40 kb)[18], although individual domains can extend much further. Such domains contain co-functioning genes that can span multiple operons; they provide improved transcriptional control, as co-functioning genes within the same supercoiling domain can become accessible for transcription together by unraveling one chromosomal coil.

Macrodomain structures in bacterial genomes might interfere with HGT[37]. It is likely that co-gained pairs cannot insert across macrodomain boundaries. While our methodology can in principle be applied to examine this prediction, we currently lack sufficient data for this, as macrodomain boundaries are currently only known in the K-12 MG1655 strain (Methods).

Bacterial mutator strains frequently arise under conditions of prolonged selection pressure[38–43], the same condition under which we expect elevated levels of HGT. We thus expect to see a correlation between HGT numbers and the presence of mutator features[44,45], i.e., deleterious mutations in genes of the methyl-directed mismatch repair pathway[44]. However, we could not detect any such mutations in the 53 analyzed genomes (Methods); a possible extension of our work might include sequenced natural mutator strains to elucidate the interaction of HGT and elevated mutation rates during adaptation.

It could be argued that using gene co-occurrence to define the basic unit of HGT might be circular: frequently co-occurring genes are by nature more likely to be jointly transferred than other gene-pairs, and co-occurrence may be a direct result of co-transfer.

However, even if SOCs were maintained by recurrent co-transfer, then this would not reduce their importance as a level of genomic organization. Indeed, the formation of SOCs might be promoted by HGT, an extension of the "selfish operon" hypothesis[19]; the examination of this hypothesis would be an interesting avenue of further research. However, we obtain identical results when defining SOCs based on shared GO annotation, confirming the functional significance of SOCs beyond their involvement in HGT.

Interestingly, an analysis of the size of co-transferred DNA segments in *E. coli* based on regional codon bias also found an upper size limit of approximately 30 kb[34]. However, the same study also found a much larger upper limit for the size of regional codon bias in *Bacillus subtilis*, extending up to approximately 180 kb[34]. One possible explanation for this disparity is that SOCs in *B. subtilis* are much larger than in *E. coli*. Furthermore, while the predominant HGT mechanism in *E. coli* is transduction, constraining the size of transferred DNA segments to carrying capacity of the phage capsid, *B. subtilis* can perform transformation and directly pick up naked DNA from the environment, which may allow uptake of much larger DNA segments. While we restricted our analysis to the acquisition and loss of genes, the analysis of regional codon bias as performed for *B. subtilis* also includes the exchange of functionally identical sequences through recombination. In such cases, natural selection on gene retention likely plays no important role in determining the fate of horizontally transferred sequences; thus, in contrast to *E. coli*, the observed size distribution of transferred DNA segments in *B. subtilis* may be determined not by natural selection favoring co-functioning gene sets, but by limits on DNA uptake via transformation.

In sum, we have shown that genes consistently co-transferred across *E. coli* strains follow a distance distribution that is consistent with SOCs, i.e., functional gene clusters that extend beyond operons; accordingly, we propose that SOCs are the basic unit of HGT. While higher-level functional clustering in bacterial genomes has been reported previously[14,32–36], these structures have so far not been linked to HGT. Future studies on the properties and the evolution of SOCs may greatly contribute to our understanding of the structure and evolutionary dynamics of prokaryotic genomes.

**METHODS**

**Reconstruction of the phylogenetic tree**

We obtained the Genbank files of 53 *E. coli* strains and 17 outgroups (Supplementary Table S1) from NCBI[46,47], extracted the amino acid sequences of the genes, and grouped them into orthologous gene families using ProteinORTHO with the synteny option[48]. This resulted in 13,138 non-orphan orthologous gene families (Supplementary Table S2: ProteinORTHO output; Supplementary Table S3: gene details).

The amino-acid sequences of the 1,334 one-to-one universal orthologs were aligned using MAFFT with default parameters[49]. We used RAxML[50] with 200 fast bootstraps and the "PROTCATAUTO" model on the concatenated alignment to estimate the phylogeny. This generated a tree with at least 60% bootstrap support at each internal branch.

Rooting at the outgroup resulted in a rooted subtree of 53 *E. coli* strains, whose 52 internal nodes are considered ancestral strains (Supplementary Fig. S1: rooted 53-strain *E. coli* tree; Supplementary Fig. S2: full unrooted tree; Supplementary Data S1: Newick format of the full tree).

**Reconstructing ancestral genomes and inferring the genes acquired through HGT**

We reconstructed the ancestral genomes using the maximum likelihood algorithm in GLOOME[25]. Based on the default parameters in the online version of GLOOME (http://gloome.tau.ac.il/), we used the option "Variable gain/loss ratio (mixture)" and "Estimate branch lengths using likelihood" (proportional to branch-specific gain/loss rates). GLOOME calculated the probability of presence of each gene at different internal nodes. If the probability of a gene was ≥0.5 at a node, then we assumed it was present, else it was absent (Supplementary Table S5: orthologous gene families in each node). We considered that a gene is lost if it is present at the start of a branch but absent at the end; a gene is gained if it is absent in a node but present in its descendant.

**Identifying the evolutionary associations between gene-pairs**

There are 16,450 orthologous gene families in 53 strains, and we represented the gain history of an orthologous gene family across the 104 branches of the tree by a vector of ones (gained) and zeros (lost) with 104 elements. Next we quantified pairwise associations of a gene family pair: we summed the occurrence of the four element-wise patterns [0,0], [0,1], [1,0] and [1,1] over the 104 rows of the vector-pair and represented the sums as a 2-by-2 contingency table. The association score of the gene family pair is defined as the decadic-logarithm of the right tail *p*-value of Fisher's exact test.

Alternatively, as GLOOME also calculates the probability to gain different genes in each branch, a possible association score for two genes can be based on the product of the two gain probabilities, summed over all branches (Supplementary Fig. S19). However, subsequent analysis showed that this approach picked up many irrelevant pairs (see the association score distribution in Supplementary Fig. S20 compared to Fig. 2).

**Null model of gene association**

The gene-pair association score we defined based on Fisher's exact test has no straight-forward statistical interpretation, because Fisher's exact test assumes independence of observations, but a gene gained on one branch cannot be gained on the subsequent branch. Hence, we estimated statistical significance from a random null model: we shuffle the presences and absences of each gene across extant strains, and then applied the same procedure to calculate the association scores.

We used the false discovery rate (FDR) to assess statistical significance. Let $N_a(t)$ and $N_n(t)$ be the number of gene-pairs with association-score more significant than $t$ ($<t$) in the empirical data and null model, respectively; a gene-pair with score $t$ has FDR$(t)=N_n(t)/N_a(t)$. Fig. 1 shows that FDR 0.05 (0.005) corresponds to an association score of -7.1289 (-9.1090).

## Five representative E. coli strains used in the analysis

We analyzed operons and autocovariances on five representative *E. coli* strains: BL21(DE3)-AM946981, O157-H7-str-Sakai, APEC-O1, IAI1, and K-12 MG1655.

## Assigning gene ontology (GO) terms to orthologous gene families

We queried the UniProt database[51] to obtain the protein entries that match our orthologous gene families. For each ortholog, we query UniProt using gene names and locus tags extracted from Genbank files. Query results with organism names not containing "*Escherichia coli*" or "*Shigella*" were removed; gene-names or locus-tags returning multiple query-entries were also removed. An ortholog could map to one or more UniProt entries, and its GO terms[30] were defined as the union of entries.

## Detecting phage association of the orthologous gene families

We can identify phage-associated genes from their GO terms and Genbank annotations. We queried the AmiGO 2 database[30] with the terms "phage" and "mobile element"; manually inspection identified 34 phage-associated GO terms (Supplementary Table S6). Further, if the word "phage" appears in the note, function, or product descriptions of a gene in its Genbank entry, its ortholog was also considered phage-associated.

We defined an orthologous gene family to be either phage-associated, not phage-associated, or with phage association uncertain. An orthologous gene family is phage-associated if it has a phage-related GO term or Genbank description. If an orthologous gene family has GO terms assigned but no evidence of phage association, then it is not phage-associated. Finally, orthologous gene families without assigned GO terms and with no evidence of phage association in the GenBank file are considered to have uncertain phage association.

## Mutual information (MI) of gene-pairs based on phylogenetic profiles

We used mutual information (MI) to infer the degree of co-occurrence of a gene family pair across multiple species. We obtained the absence/presence-profile of 3069 genes across 233 γ-proteobacterial species (including *E. coli*) from Table S14 of *Babu et al.*[52]; the profile vector of a gene has 233 elements, which can either be 0 (absent) or 1 (present). The occurrence of gene families A and B on a genome can be one of the four patterns (1) [0,0], (2) [1,0], (3) [0,1] and (4) [1,1]. We denoted their probabilities as $p_1$, $p_2$, $p_3$, $p_4$, where the probability of presence for A and B are $q_A=p_2+p_4$ and $q_B=p_3+p_4$. The information entropy of gene A, $H(A)$, is

$$H(A) = -q_A \log_2 q_A - (1-q_A)\log_2(1-q_A) \quad \quad 1$$

The information entropy for the joint distribution of A and B, $H(A,B)$, is

$$H(A,B) = -\sum_{n=1}^{4} p_n \log_2 p_n \quad \quad 2$$

Finally, their MI, $I(A:B)$, is

$$I(A:B) = H(A) + H(B) - H(A,B) \quad \quad 3$$

The MI of a gene family pair is bounded between 0 and 1, with 0 indicating no relation and 1 indicating perfect co-occurrence.

**Delineation of SOCs through autocovariance**

We used autocovariance (AC), denoted as $G(x)$, to measure the degree of clustering of (i) co-occurring (CO-AC) and (ii) functionally coupled (GO-AC) gene families. Given a nucleotide at site $i=0$ in gene A, $G(x)$ measures the probability for another site at genomic position $j$ ($x=j-i$) in gene B, such that A and B are considered to be (i) co-occurring if A and B have an MI above a specified cutoff value, or (ii) co-functioning if the number of their common GO terms[30] is above a specified cutoff. Let $g_i(x)$ be a discrete function that maps distance $x$ to ones and zeros: $i$ can be any nucleic site on the genome within a gene; $x$ is a positive integer. We define $g_i(x)=1$ if (i) $x$ and $x+1$ are sites of two different genes that are co-occurring (or co-functioning), (ii) both genes are not associated with phages, and (iii) both do not overlap with mobile element regions indicated in the Genbank file; otherwise $g_i(x)=0$. AC is then defined as

$$G(x) = \frac{1}{n}\sum_{i \in N} g_i(x) \quad \quad 4$$

where $N$ is the set of all $n$ nucleic sites in the genes considered.

We expect a gene-pair within the same SOC to have a higher MI than random gene-pairs. Thus, we expect $G(x)$ to have the form

$$G(x) = G_{cluster}(x) + G_0 \quad \quad 5$$

where $G_{cluster}(x)$ is the part that reflects the distance between gene-pairs within a cluster, while $G_0$ is the background value. Gene-pairs with small $x$ are likely to be from the same cluster; but as $x$ increases, this chance decays to $G_0$. We used a sliding frame of 10 bp to reduce computation times.

CO-AC and GO-AC have different $G_0$. To make them comparable, we rescaled $G(x)$ into $\widetilde{G}(x)$ as

$$\tilde{G}(x) \quad \frac{G(x) - \langle G(x) \rangle}{\max(G(x)) - \langle G(x) \rangle}$$

where $\langle G(x) \rangle$ is the mean of $G(x)$, and $\max(G(x))$ is the maximum of $G(x)$. This definition causes $\tilde{G}(x)$ to peak at 1, whereafter it decays to 0. We log-binned $\tilde{G}(x)$ to remove noise, and cut it off at the point where it first crosses the x-axis at large $x$. Ignoring nucleotide pairs from within the same gene leads to noise at low distances, which we also cut off. Subsequent normalization of $\tilde{G}(x)$ converts it into an estimation of the gene-pair distance distribution in co-occurring or co-functioning gene clusters, which is an approximation to $G_{cluster}(x)$.

We tested this approximation by applying it to gene-pairs within *E. coli* operons. Supplementary Fig. S12 compares the genomic distance distributions of gene-pairs in operons with the distributions approximated using the rescaled and normalized GO-AC with gene-pairs located in different operons ignored. It reveals that while the bulk of the approximate distribution is slightly biased to the right, the right tail is well approximated.

## Search for mutator strains

Deletion or damage of the coding sequence of any of the four methyl-directed mismatch repair (MMR) genes reported in *LeClerc et al*[44] (mutH, mutL, mutS, uvrD) indicates a mutator strain. As these four genes are present across all 53 strains, we investigated their integrity. For each of the four, we aligned the amino acid sequences of the 53 alleles using MAFFT[49]; we considered the divergence of all allele pairs, and found that the majority of them have divergence <1.5%, and the most divergent pair is <3.5% (Supplementary Fig. S21). We found no evidence of frameshifts, nonsense mutations, or large indels rendering an allele nonfunctional; we thus could not identify any mutator strains in our dataset.

## Overlap of HGT clusters with macrodomains

We examined our dataset to check if any of the co-gained pairs span across macrodomain boundaries[37,53]. There are multiple attempts to measure the start and end of the macrodomain boundaries in K-12 strains[37,54,55]; the results of these measurements are listed in *Messerschmidt et al.*[53]. In total, we found 534 non-phage-associated orthologous co-gained orthologous family pairs at FDR 0.05, which appeared 5,273 times across the 53 genomes. However, only one of these pairs occurs in the K-12 MG1655 strain, with the two genes separated by a distance of 1.3 Mb; this is likely a false positive, as it sits in the long tail of the distribution in Fig. 2. When we also include phage-associated genes, we obtained a total of 16,566 pairs across the 53 genomes. 48 of them are in K-12 MG1655, with another likely false positive pair separated by 0.9 Mb. The remaining 46 pairs can be placed in three 30 kb clusters, none of which crosses a macrodomain boundary. However, due to the small sample size (three clusters), we cannot conclude whether HGT clusters indeed avoid macrodomain boundaries.

# ACKNOWLEDGEMENTS

This work was supported by the German Research Foundation (DFG-grant CRC 680 to MJL).

# AUTHOR CONTRIBUTIONS

Both authors designed and conducted the research. TYP prepared all figures, and both authors wrote the manuscript.

# COMPETING FINANCIAL INTERESTS

The authors declare no competing financial interests.

# FIGURE LEGENDS

**Figure 1. Cumulative distribution function of pairwise gene association scores**
Many gene-pairs show much stronger co-gain scores than expected from random HGT. Distribution of the score for pairwise gene associations $t$ in the empirical data (solid line) and the null model (dashed line), based on the maximum-likelihood ancestral genome reconstructions. The two vertical dotted lines at $t$=-7.1289 and -9.1090 correspond to FDRs of 0.05 and 0.005, respectively. See Supplementary Figure S3 for the results obtained from the high-confidence HGT data set. The pairs of consistently co-gained genes at FDR 0.05 are available as Supplementary Table S4.

**Figure 2. Cumulative distance distribution function of co-gained gene family pairs**
The distance distribution of co-gained gene-pairs drops sharply at around 30 kb. Cumulative distribution function of genomic distances between all co-gained gene-pairs (black lines) and between co-gained gene-pairs that are known to be non-phage-associated (brown lines), at FDR=0.05 (solid lines) and FDR=0.005 (dashed lines). The distributions of all co-gained gene-pairs show a tail that extends to 2.5Mb, while the distributions for non-phage-associated gene-pairs are cut off at approximately 30 kb. See Supplementary Fig. S5 for corresponding probability density functions, and Supplementary Figure S4 for the results obtained from the high-confidence HGT data set.

**Figure 3. Cumulative distance distribution of co-gained gene-pairs compared to that of genes in operons and in SOCs**
The observed cumulative distance distribution is consistent with distances within SOCs, but not within operons. Brown solid line: observed cumulative distance distribution of co-gained gene-pairs at FDR=0.05. Black dashed line: cumulative pairwise distance distribution of genes in operons[31], averaged over five *E. coli* strains. Black dash-dot line: SOCs based on GO-AC curves requiring at least one common GO term (*n*=1), averaged over five strains. Black solid line: SOCs based on CO-AC of five *E. coli* strains, cut off at an MI value of 0.0001. See Supplementary Fig. S10 for the operon distributions of individual strains; S11 for distributions of the CO-AC at different MI cutoffs; and S13 for distributions of GO-AC (*n*=1) for the five strains.

## TABLES

**Table 1. Statistics of co-transferred gene-pairs**

|  | Number of pairs[1] | | | | |
|---|---|---|---|---|---|
|  | total | both annotated in GO | both not phage-associated | distance < 30 kb | distance ≥ 30 kb |
| FDR 0.05 | 1790 | 598 | 534 | 518 | 16 |
| FDR 0.005 | 949 | 478 | 454 | 444 | 10 |

[1] horizontal lines starting at one column mean that all further columns underneath are subsets of this column